\documentclass[aps,twocolumn,showpacs,amsmath,amssymb,superscriptaddress,longbibliography,prl]{revtex4-2}

\usepackage{color,soul}

\usepackage{float}
\usepackage{graphicx}
\usepackage{svg}
\usepackage{longtable}
\usepackage{amsmath}
\usepackage{xfrac}
\usepackage{multirow}
\usepackage{pifont}

\usepackage[
    colorlinks,
    linkcolor={blue!80!black},
    citecolor={blue!80!black},
    urlcolor={blue!80!black}
]{hyperref}

\begin{document}

\title{Ferroelectric $p$-wave magnets}

\author{Jan Priessnitz}
\affiliation{Max Planck Institute for the Physics of Complex Systems, Nöthnitzer Str. 38, 01187 Dresden, Germany}

\author{Anna Birk Hellenes}
\affiliation{Institute of Physics, Czech Academy of Sciences, Cukrovarnick\'a 10, 162 00, Praha 6, Czech Republic}
\affiliation{Institut f\"ur Physik, Johannes Gutenberg Universit\"at Mainz, D-55099 Mainz, Germany}

\author{Riccardo Comin}
\affiliation{Department of Physics, Massachusetts Institute of Technology, Cambridge, 02139, Massachusetts, USA}

\author{Libor Šmejkal}
\affiliation{Max Planck Institute for the Physics of Complex Systems, Nöthnitzer Str. 38, 01187 Dresden, Germany}
\affiliation{Max Planck Institute for Chemical Physics of Solids, Nöthnitzer Str. 40, 01187 Dresden, Germany}
\affiliation{Institute of Physics, Czech Academy of Sciences, Cukrovarnická 10, 162 00 Praha 6, Czech Republic}

\date{\today}

\begin{abstract}
Couplings between ferroelectric and magnetic orders offer promising routes toward low-dissipation electronics. However, such couplings are notably rare, largely due to the poor compatibility between insulating band structures and ferromagnetism. Here, we study a different strategy: we identify previously overlooked time-reversal-symmetric $p$- and $f$-wave spin-polarized insulating electronic states in ferroelectrics with noncollinear magnetic sublattices. We show that combining spin and magnetic group theory enables a systematic classification of the origin of polar symmetry breaking.  We distinguish crystallographic, exchange-, or spin–orbit-driven mechanisms. Furthermore, we identify more than 50 candidate materials. Using first-principles calculations, we demonstrate a pristine, time-reversal-symmetric $p$-wave spin-polarized electronic structure in the well-known multiferroic $\mathrm{GdMn_2O_5}$. We further show that its $p$-wave order can be switched electrically, opening alternative paths toward spintronic and multiferroic functionalities in this class of materials.
\end{abstract}

\maketitle

\textit{Introduction.}--- Multiferroics are materials that combine multiple spontaneous orderings ~\cite{Spaldin2010,spaldin_advances_2019}. Multiferroics combining ferroelectricity and magnetic ordering are commonly classified into two categories, based on the origin of the ferroelectric ordering~\cite{khomskii_classifying_2009}.
In Type-I multiferroics, magnetism and ferroelectricity have distinct origins, with the latter being induced by structural distortions in the crystal and not being directly dependent on the magnetic order. In contrast, Type-II
multiferroics exhibit ferroelectricity induced by the spin order, which lowers the symmetry of the crystal~\cite{Tokura_2014}. Here, the nonrelativistic and spin-chirality-driven ferroelectricity~\cite{PhysRevLett.95.087206}, and magnetoelectric coupling~\cite{kimura_magnetic_2003}  can be very strong. However, both types require
an insulating band structure, which is generally poorly compatible
with ferromagnetism~\cite{spaldin_advances_2019}.

\begin{figure}[t]
  \includegraphics[width=\columnwidth]{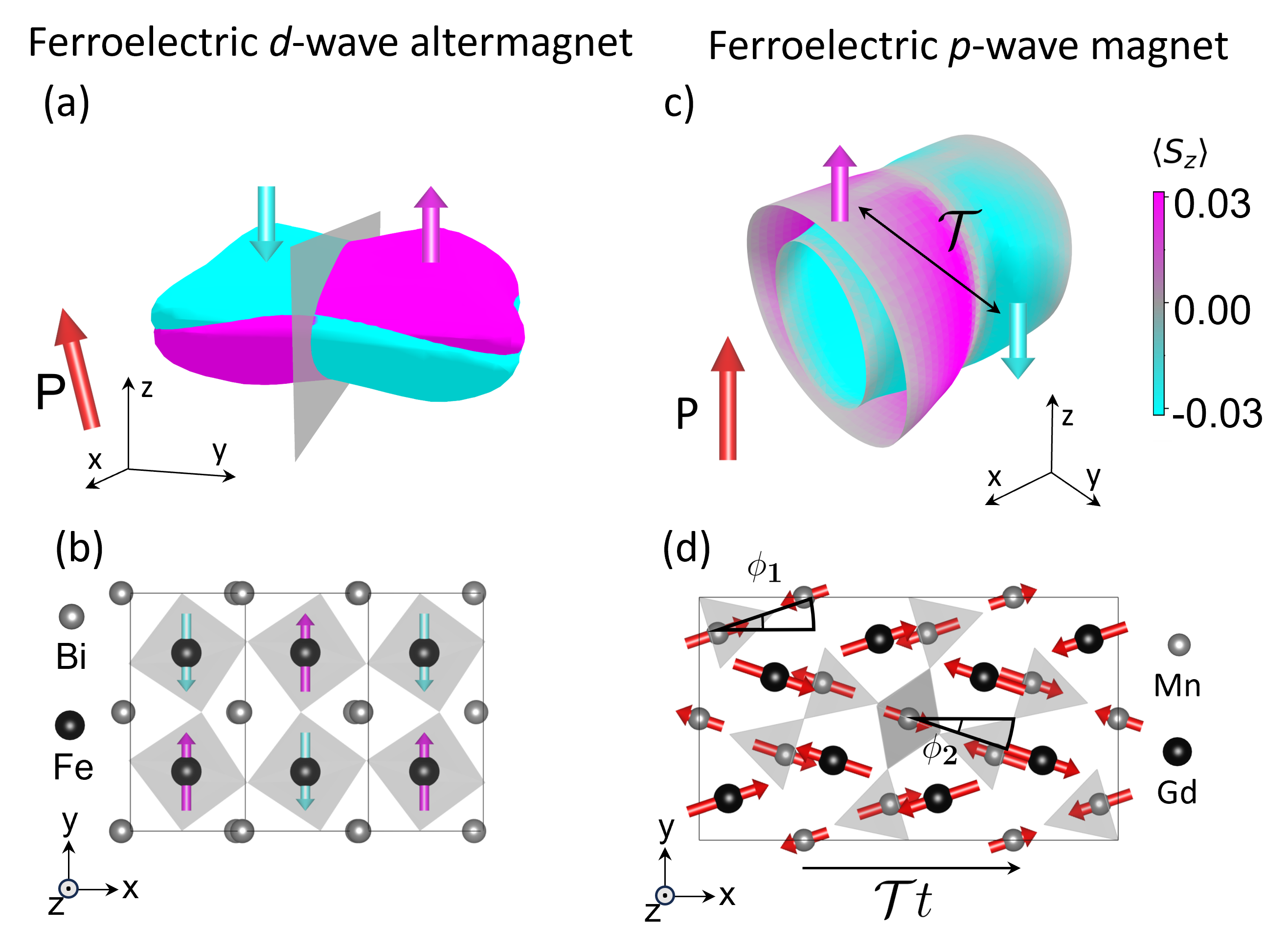}
  \centering
  \caption{(a) Altermagnetic $d$-wave spin polarization with highlighted nodal plane on top of constant-energy isosurface at energy $0.5~\mathrm{eV}$ below the Fermi level, and (b) Magnetic structure of ultrathin $\mathrm{BiFeO_3}$~\cite{fratian2026topologicaltexturesemergentaltermagnetic}.  
  (c) Time-reversal $\mathcal{T}$ symmetric $p$-wave spin polarization on top of constant-energy isosurface at energy $ 0.1~\mathrm{eV}$ below the Fermi level calculated for $\mathrm{GdMn_2O_5}$ with a ferroelectric polarization $P$ along the $z$-axis. Color indicates the $z$-component of the spin polarization.  (d) Coplanar magnetic structure of $\mathrm{GdMn_2O_5}$. O atoms are located on the vertices of the grey-shaded polyhedra.}
  \label{fig:BiFeO-GdMnO}
\end{figure}

An alternative strategy to combine ferroelectricity, or antiferroelectricity, with spontaneous spin polarization in altermagnets has been suggested more recently~\cite{smejkal2024altermagneticmultiferroicsaltermagnetoelectriceffect, PhysRevLett.134.106802,PhysRevLett.134.106801}.
Altermagnets are delimited as a distinct type of collinear magnets in addition to conventional ferromagnets and antiferromagnets via spin group theory~\cite{smejkal_beyond_2022}. They exhibit time-reversal-symmetry breaking nonrelativistic  $d$-, $g$-, or $i$-wave spin-polarized order~\cite{smejkal_crystal_2020, smejkal_beyond_2022, doi:10.1073/pnas.2108924118} in both direct and reciprocal space~\cite{bhowal_ferroically_2024, jaeschkeubiergo2025atomicaltermagnetism}. They were identified in both bulk~\cite{smejkal_beyond_2022,smejkal_emerging_2022} and monolayer materials~\cite{doi:10.1021/acs.nanolett.5c02121,  mazin2023inducedmonolayeraltermagnetismmnpsse3}.
Recently, altermagnetic order has been proposed~\cite{smejkal2024altermagneticmultiferroicsaltermagnetoelectriceffect,v3fg-6smc} and experimentally indicated in the ultrathin variant of the known multiferroic $\mathrm{BiFeO_3}$~\cite{fratian2026topologicaltexturesemergentaltermagnetic}. Its $d$-wave altermagnetic order~\cite{smejkal2024altermagneticmultiferroicsaltermagnetoelectriceffect,fratian2026topologicaltexturesemergentaltermagnetic} is shown in \autoref{fig:BiFeO-GdMnO}(a).

\begin{table*}[th!]
\centering
\begin{tabular}{|c|c|c|c|}
\hline
 \textbf{polar $\mathbf{p}$-wave magnets} & $\mathrm{Ni_2Mo_3O_8}$ & $\mathrm{GdMn_2O_5}$ & $\mathrm{CeNiAsO}$ \\
 \includegraphics[width=0.32\columnwidth]{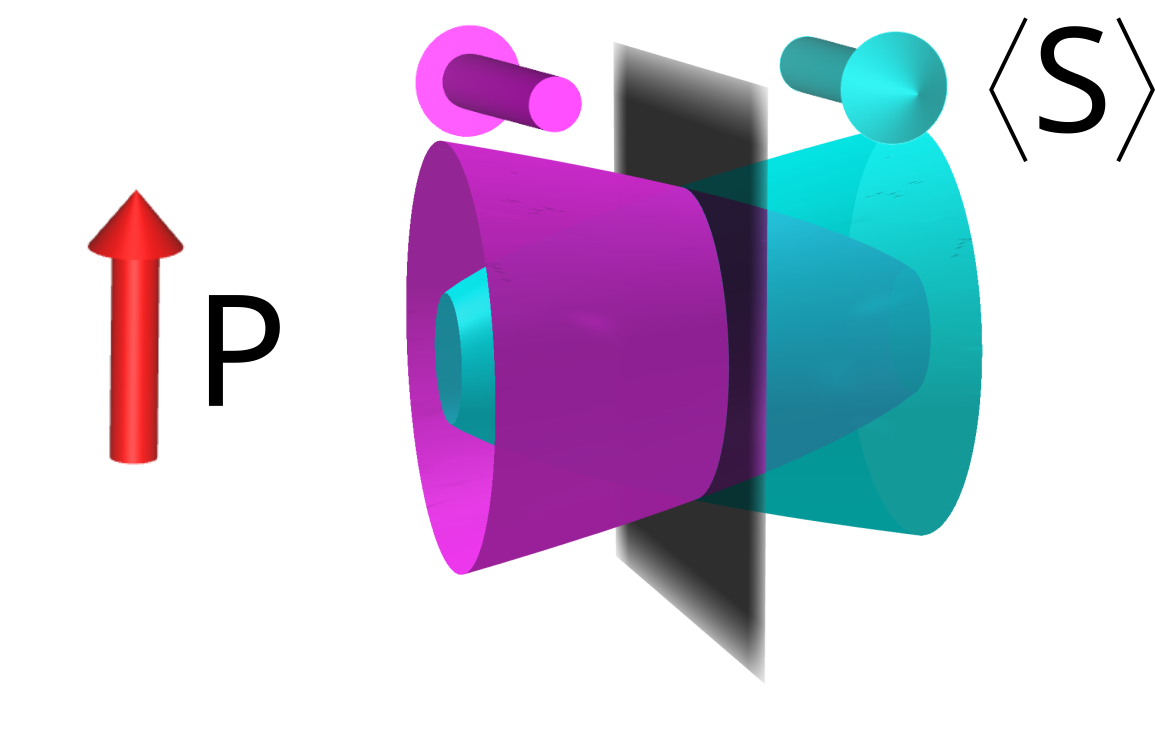} & \includegraphics[width=0.5\columnwidth]{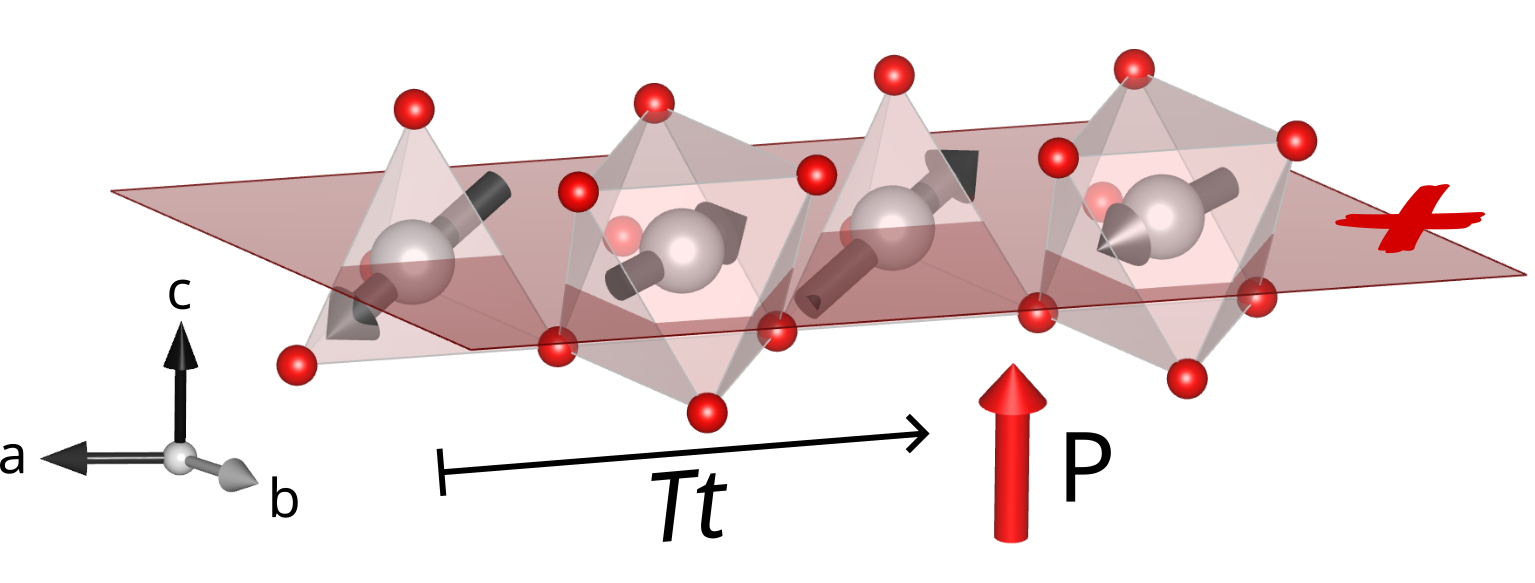} & \includegraphics[width=0.5\columnwidth]{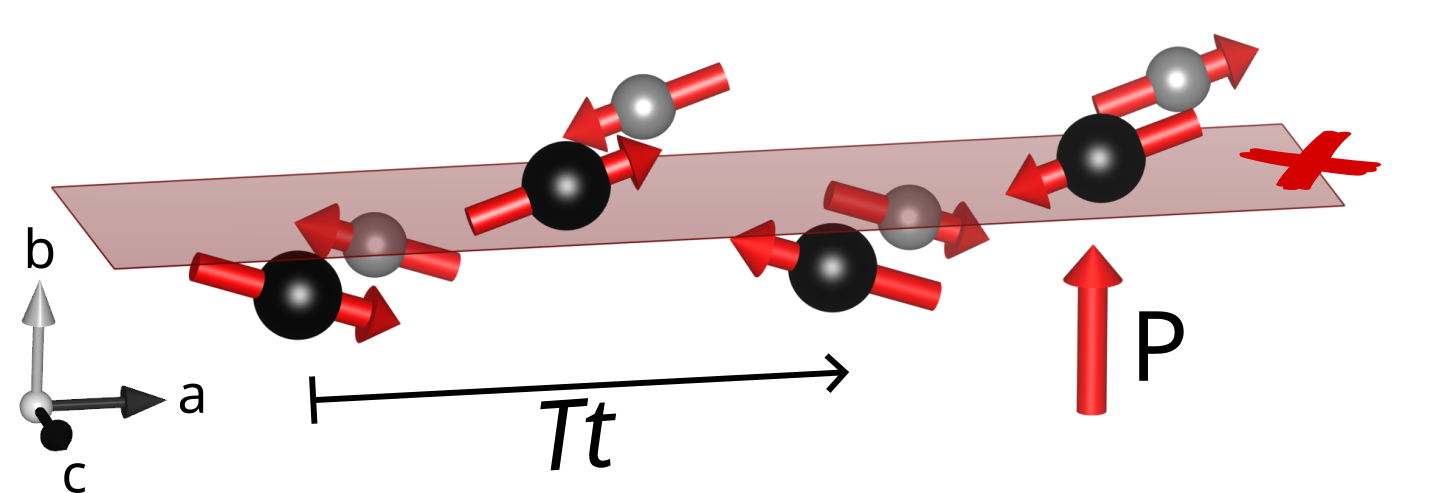} & \includegraphics[width=0.5\columnwidth]{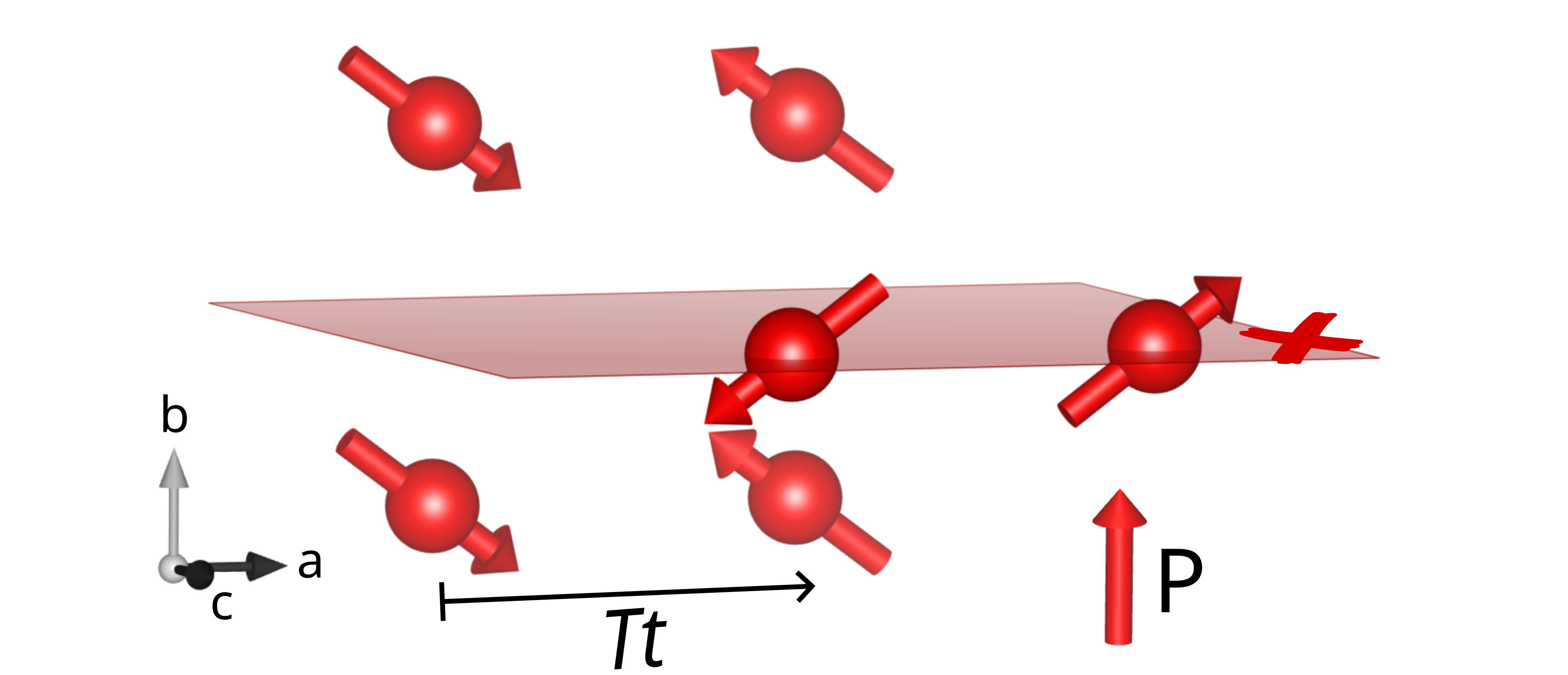} \\ \hline
class & Type-I & Type-IIa & Type-IIb \\ \hline \hline 

$G_{NM}$/$G_S$/$G_M$ polar? & $\checkmark$ / $\checkmark$ / $\checkmark$ & \ding{55} / $\checkmark$ / $\checkmark$ & \ding{55} / \ding{55} / $\checkmark$ \\ \hline
\# of $p$-/$f$-/$h$-wave candidates & 10 / 2 / 0 & 27 / 1 / 0 & 2 / 10 / 0 \\ \hline

  \end{tabular}
    \caption{Classification table of polar odd-parity-wave magnets with material examples, symmetry criteria and known material counts with $p$-, $f$- or $h$-wave spin splitting types. $G_{NM}$, $G_{S}$ and $G_{M}$ denote the nonmagnetic point group, and lattice parts of the spin and magnetic point group, respectively. }
  \label{tbl:classification}
\end{table*}

In this work, we address the question of whether ferroelectricity can be combined with nonrelativistic time-reversal-symmetric spin splitting, identified in a recently proposed class of noncollinear odd-parity-wave magnets~\cite{hellenes2024pwavemagnets, JUNGWIRTH2025100162, Chakraborty2025}.
Odd-parity-wave magnets exhibit $p$-, $f$-, or $h$-wave spin polarization~\cite{hellenes2024pwavemagnets, JUNGWIRTH2025100162, PhysRevLett.133.236703, zk69-k6b2}.
Experimental evidence of $p$-wave spin polarization has recently been demonstrated in an incommensurate magnet lacking the archetypal $\mathcal{T}t$ symmetry, namely the multiferroic insulator $\mathrm{NiI_2}$~\cite{Song2025-uf}. Importantly, the $p$-wave spin polarization in $\mathrm{NiI_2}$ can be switched electrically~\cite{Song2025-uf,Li2026}. Furthermore, anisotropic transport consistent with $p$-wave spin polarization, along with a large anomalous Hall effect arising from additional symmetry breaking, has been reported in metallic alloys $\mathrm{Gd_3(Ru_{1-\delta}Rh_{\delta})_{4}Al_{12}}$ that also lack $\mathcal{T}t$ symmetry~\cite{Yamada2025}.

Here, we demonstrate the possibility of combining pristine commensurate and \textit{time-reversal-symmetric} odd-parity-wave magnetism with ferroelectricity. In \autoref{fig:BiFeO-GdMnO}(c-d), we identify a material candidate, a well-known multiferroic $\mathrm{GdMn_2O_5}$. Unlike in the case of ferroelectric altermagnets with time-reversal-broken spin polarization parallel to the direct space spin order, our ferroelectric odd-parity-wave spin-polarized multiferroics exhibit collinear and time-reversal-symmetric ``antialtermagnetic''~\cite{JUNGWIRTH2025100162, jungwirth_symmetry_2026,Mitscherling2026} spin polarization, $\langle S_{\perp} (\boldsymbol{k}) \rangle = \langle S_{\perp} (-\boldsymbol{k}) \rangle$, as illustrated in \autoref{fig:BiFeO-GdMnO}(c-d), that is perpendicular to the direct space order~\cite{hellenes2024pwavemagnets}. In the case of $\mathrm{GdMn_2O_5}$, the collinear time-reversal-symmetric spin splitting is a consequence of the coplanar magnetic ordering and the $\mathcal{T}t$ symmetry, shown in \autoref{fig:BiFeO-GdMnO}(d). Detailed spin symmetry analysis of $p$-wave magnets is discussed in Ref.~\cite{hellenes2024pwavemagnets}.

We identify 52 ferroelectric, odd-parity-wave-magnetic material candidates, and classify them as Type-I and II multiferroics and according to their partial-wave character ($p$-, $f$-, or $h$-wave). Remarkably, Type-II multiferroics can be further split into two branches depending on whether relativistic spin-orbit coupling (SOC) is necessary to induce polar order. In Type-IIa (dubbed ``nonrelativistic''), it is not necessary, while in Type-IIb (dubbed ``relativistic''), it is.
Furthermore, we demonstrate an unconventional $p$-wave-magnetoelectric coupling in $\mathrm{GdMn_2O_5}$~\cite{lee_giant_2013,ponet_topologically_2022}. We show that this coupling can be used to electrically switch the $p$-wave spin polarization and can be used in an electrically controlled memory device.

\textit{Symmetry classification of polar odd-parity wave magnets into Type-I, IIa, and IIb.}--- 
We now delimit the material class of polar time-reversal symmetric odd-parity-wave magnets via spin symmetries. Such materials must fulfill separate criteria for both ferroelectric and $p$-wave order. On one hand, for the odd-parity-wave magnetism, the combination of spatial-inversion and time-reversal symmetry $\mathcal{PT}$, as well as spatial-inversion symmetry $\mathcal{P}$ alone, must be broken. Furthermore, we focus here on the archetypal $p$-wave magnets where the combination of time-reversal and translation $\mathcal{T}t$ must be present~\cite{hellenes2024pwavemagnets}.
On the other hand, the polar order can only arise in materials belonging to a polar symmetry group which leaves at least one spatial direction invariant. This is a stronger requirement than only broken inversion symmetry.

In line with the classification of conventional multiferroics~\cite{khomskii_classifying_2009}, we classify polar odd-parity-wave magnets based on the microscopic origin of the symmetry breaking leading to the polar order. Moreover, with the help of spin group theory, we further split the Type-II category into nonrelativistic Type-IIa and relativistic Type-IIb. Our classification is summarized in \autoref{tbl:classification} with material examples, symmetry criteria, and known material candidate counts. 

Type-I class labels a polar order that arises directly from the crystal structural distortions represented in the polar nonmagnetic point group $G_{NM}$. In the example structure of $\mathrm{Ni_2Mo_3O_8}$, shown in \autoref{tbl:classification}, the non-equivalent Ni atoms positioned above and below the red $M_z$ mirror plane clearly break this mirror symmetry, yielding a polar nonmagnetic point group.

In stark contrast, Type-IIa labels structures with non-polar $G_{NM}$, in which polar symmetry breaking occurs once magnetic order is established. The symmetry of the magnetic ordering without considering SOC is described by nonrelativistic spin group theory~\cite{litvin_spin_1974, smejkal_beyond_2022, jungwirth_symmetry_2026, PhysRevX.12.021016} which considers pairs of generally distinct operations in spin and lattice space. The required polar symmetry breaking manifests in the lattice part~\cite{jungwirth_symmetry_2026} (also called orbital part~\cite{PhysRevB.111.085147}) of the spin  group $G_S = \left\{ R_o: [R_s || R_o] \in \mathcal{G} \right\}$, where $\mathcal{G}$ is the full spin group. Its elements in the form $[R_s || R_o]$ are spin symmetry operations combining operation $R_s$ acting in the spin space and operation $R_o$ acting in the lattice space.
In the example of $\mathrm{GdMn_2O_5}$, the polar structure is induced by the magnetic ordering, which breaks the glide plane symmetry $\left\lbrace M_y | t({\sfrac{1}{4}, 0, 0}) \right\rbrace$ present in the nonmagnetic state.

Magnets in the Type-IIb category become polar only after taking relativistic SOC into account. This is reflected in the lattice part of the magnetic group $G_M$, while $G_S$ must remain non-polar. In the example of $\mathrm{CeNiAsO}$, Type-IIb polar ordering can be interpreted in terms of the relative orientation of the spin frame of reference with respect to the unit cell. If it were rotated such that the coplanar magnetic ordering would lie in the $ac$-plane (instead of the $ab$-plane), the mirror indicated in \autoref{tbl:classification} would cease to be broken and the material would become non-polar.
 
The hierarchy of the symmetry breaking is determined by the relation $G_{M} \subseteq G_{S} \subseteq G_{NM}$~\cite{Etxebarria:cam5007} and the fact that a subgroup of a polar group is always polar, too. Remarkably, some materials can have multiple polar symmetry breakings at different levels of the symmetry breaking hierarchy. Consequently, the breakings can gradually unlock additional degrees of freedom in the electric polarization vector. In these cases, we classify such materials based on the first symmetry breaking in the hierarchy. An example of this is $\mathrm{Cu_2MnSi_4}$, which is further discussed in Supplemental Information~\cite{supp}.

\textit{Material candidates.}---
We conducted a material search through the \textsc{Magndata} database of experimentally reported magnetic structures~\cite{Gallego:ks5532} and found 52 material candidates. \autoref{tbl:classification} includes the material counts in the symmetry categories of Type-I, IIa, and IIb, and partial-wave order ($p$-, $f$- or $h$-wave). Interestingly, the majority of the materials (40) belong to the Type-II category. This is in stark contrast to altermagnetic multiferroics, where only a single Type-II material candidate has been identified~\cite{smejkal2024altermagneticmultiferroicsaltermagnetoelectriceffect}.  Classification and full list of the materials are available in the Supplemental Information~\cite{supp}.

\begin{figure}[t]
  \includegraphics[width=\columnwidth]{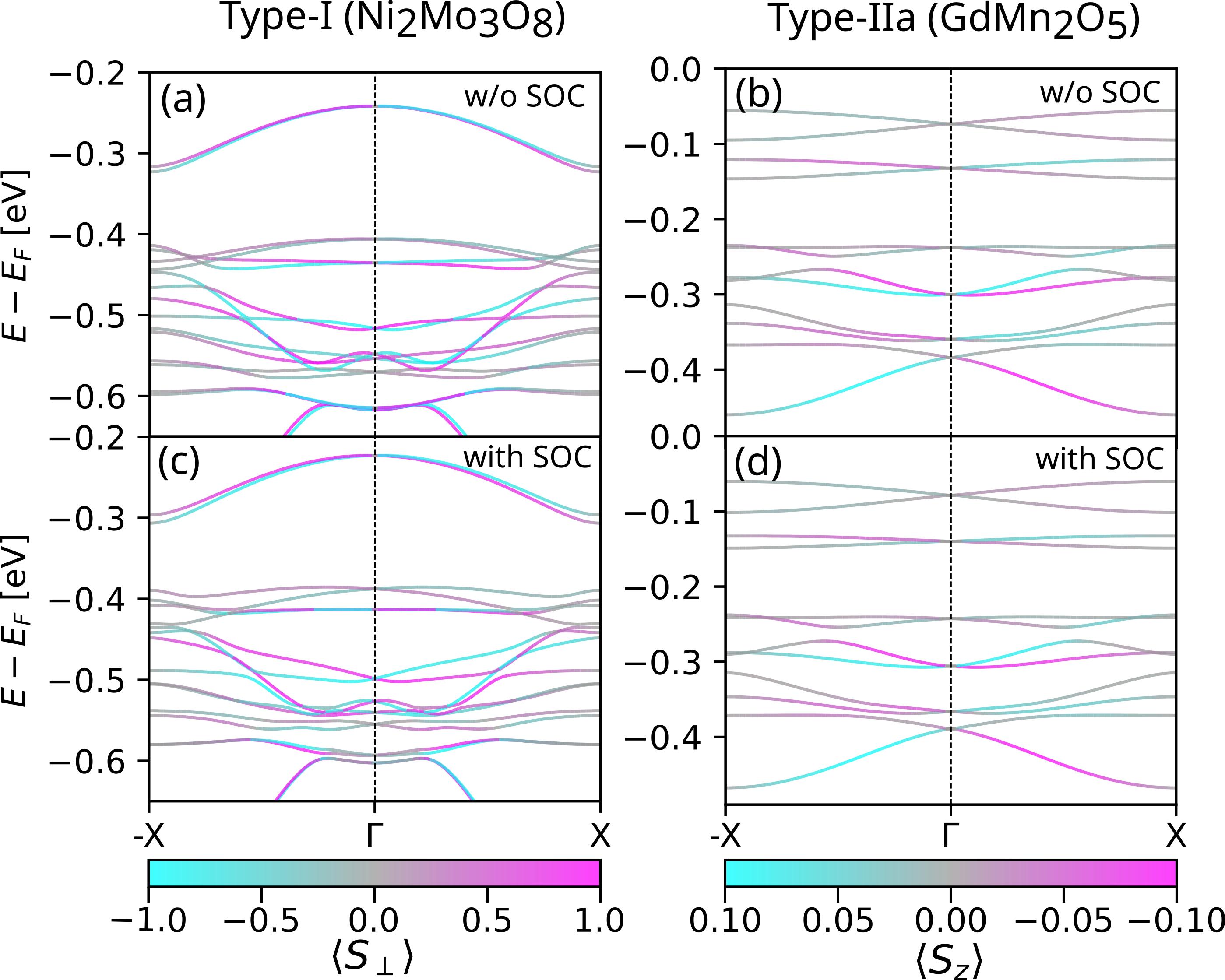}
  \centering
  \caption{ Band structures showing $p$-wave spin splitting in Type-I multiferroic $\mathrm{Ni_2Mo_3O_8}$ and Type-II multiferroic $\mathrm{GdMn_2O_5}$, without and with SOC. High-symmetry points are located at $\Gamma = (0, 0, 0)$, $X = (\pi/a, 0, 0)$.}
  \label{fig:multiferroic-types}
\end{figure}

\textit{First-principles calculations of $p$-wave spin polarization in $Ni_2Mo_3O_8$ and $GdMn_2O_5$.}---According to our symmetry analysis, pyroelectric $\mathrm{Ni_2Mo_3O_8}$~\cite{PhysRevResearch.5.033099} and ferroelectric $\mathrm{GdMn_2O_5}$~\cite{ponet_topologically_2022} both belong to the class of odd-parity-wave polar magnets. We present first-principles calculations of band structures of the two materials in \autoref{fig:multiferroic-types}, showing large nonrelativistic $p$-wave spin splitting. By comparing the calculations without SOC, shown in panels (a,b), and with SOC, shown in panels (c,d), we see that the origin of the spin splitting is largely nonrelativistic and the role of the relativistic SOC is minor.

Interestingly, the strength of the $p$-wave spin polarization is nearly quantized in $\mathrm{Ni_2Mo_3O_8}$, while it is much weaker in $\mathrm{GdMn_2O_5}$. That is consistent with our model calculations, which demonstrate the proportionality of the spin polarization strength to the ratio of exchange-dependent to conventional electron hopping~\cite{khodas2026nonrelativisticisingsuperconductivitypwavemagnets,Mitscherling2026}. More detailed band structures of $\mathrm{GdMn_2O_5}$ are available in the Supplemental Information~\cite{supp}.

\textit{$P$-wave-magnetoelectric coupling in $GdMn_2O_5$.}---
Next, we examine the Type-IIa multiferroic $\mathrm{GdMn_2O_5}$, introduced in \autoref{fig:BiFeO-GdMnO}(c,d), in greater detail. 
We demonstrate that a coupling of the $p$-wave-magnetic and ferroelectric order, a $p$-wave-magnetoelectric coupling (termed antialtermagnetoelectric), is present in this material. This coupling is highly relevant in the context of electric switching of the $p$-wave magnetic order.

We assume an experimentally reported magnetic ground state listed in \textsc{Magndata} (label~1.54)~\cite{lee_giant_2013}. The magnetic structure is coplanar -- all moments lie in the $xy$-plane. Moments are tilted from the $x$-axis either by $\phi_1 = \phi_0 = 18.5^{\circ}$ or $\phi_2 = - \phi_0 = - 18.5^{\circ}$.
Its ferroelectric properties, and connection to magnetism, have been studied extensively~\cite{dai_multiple-valued_2020,tsujino_magnetoelectric_1992,bukhari_magnetoelectric_2016,lee_giant_2013,yahia_experimental_2018,ponet_topologically_2022,yin_pressure-induced_2016, PhysRevLett.134.016708}, as well as ferroelectric properties of other materials in the $\mathrm{RMn_2O_5}$ family~\cite{blake_spin_2005,kaddar_theoretical_2021,wang_first-principles_2008,el_hallani_first-principles_2013,wang_ferroelectricity_2007,hur_colossal_2004,higashiyama_magnetic-field-induced_2005}.

The crystal structure belongs to space group $Pbam$ (\#55) and point group $mmm$, which is centrosymmetric and non-polar. Thus, magnetic order is necessary to induce polar order.
The magnetic structure, when not considering SOC, is described by the spin space group, which can be decomposed as a combination of a direct and semidirect product $\mathcal{G} = \mathcal{G}_\mathrm{SO} \times (\mathcal{G}_\mathrm{ST} \rtimes \mathcal{G}_\mathrm{NT})$. Here $\mathcal{G}_\mathrm{SO} = \left \{ [E || E], [M_z || E] \right \}$ represents the spin-only group of the coplanar magnetic order, $\mathcal{G}_\mathrm{ST} = \{ [E || E], [C_{2z} || t(\sfrac{1}{2}, 0, 0)] \}$ the spin-translation group, and $\mathcal{G}_\mathrm{NT} = \{ [E || E]$, $[E || M_{z}]$, $[M_{x} || M_{x}| t(\sfrac{1}{4}, \sfrac{1}{2}, 0)]$, $[M_x || C_{2y}| t(\sfrac{1}{4}, \sfrac{1}{2}, 0)] \}$ the nontrivial spin group. $\mathcal{G}$ does not contain any orbital-space mirror symmetry $[R_s || M_y]$ and therefore allows electric polarization in the form $\vec{P} = (0, P_y, 0)$.
The time-reversal-symmetric, collinear out-of-plane spin splitting is enforced by $[C_{2z} || t(\sfrac{1}{2}, 0, 0)]$, a combination of $\mathcal{T}t$ symmetry and coplanar symmetry $[M_{z} || E]$.  

\begin{figure}
    \centering
    \includegraphics[width=\columnwidth]{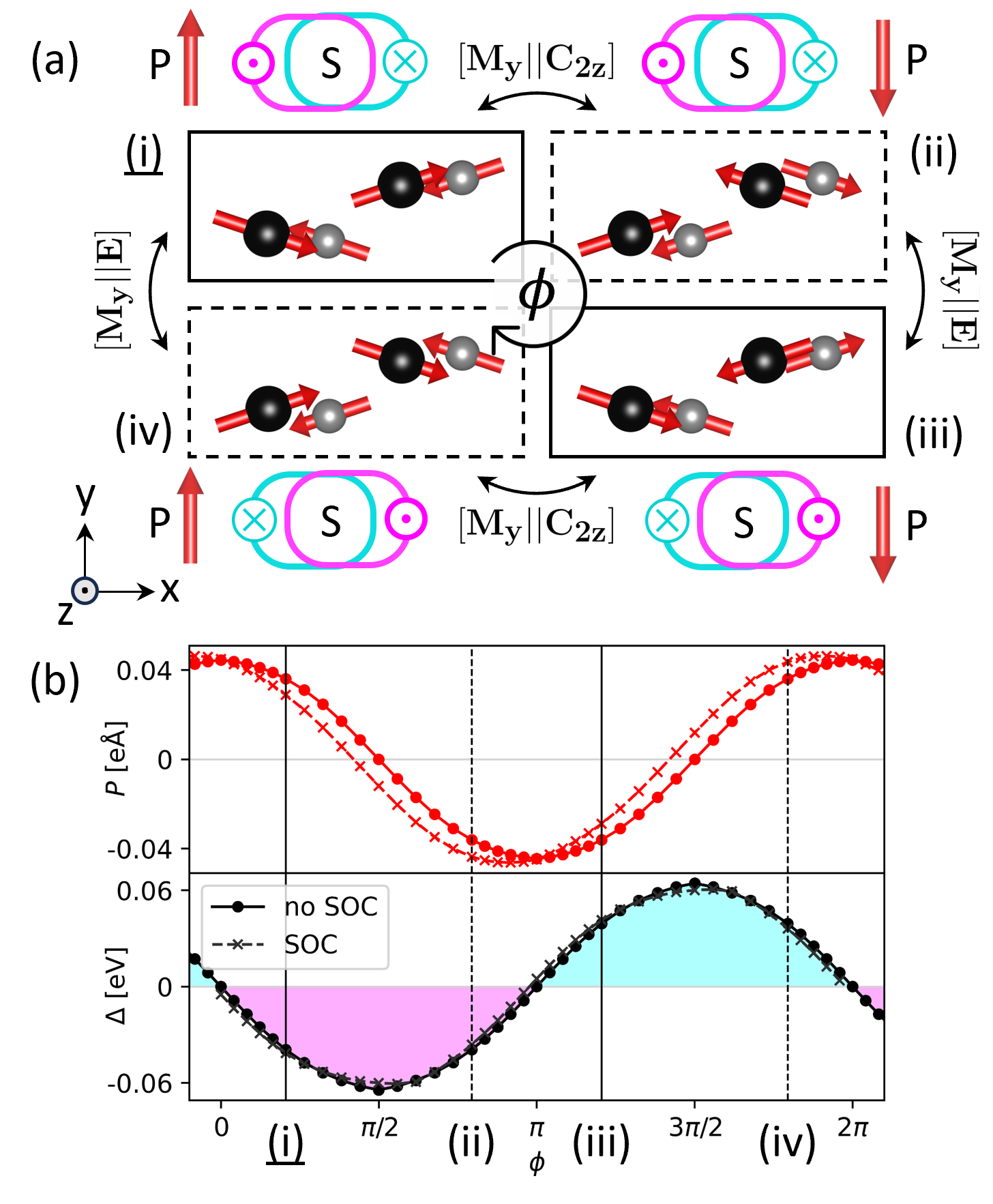}
    \caption{\textbf{Ferroelectric switching of $p$-wave magnetism: $p$-wave-magnetoelectric coupling.} (a) Four coplanar magnetic structures of $\mathrm{GdMn_2O_5}$ (only parts of the unit cell displayed) with combinations of spin (S) and electric (P) polarization signs, energetically degenerate without SOC. SOC splits the four states into two pairs of degenerate states -- states (i) and (iii) become ground states. State (i) is experimentally reported in \textsc{MAGNDATA}. (b) First-principles-calculated dependence of electric polarization from the electrons $P$ and spin splitting strength $\Delta$ of the valence band at high-symmetry point $X = (\pi/a, 0, 0)$ vs. relative rotation of the magnetic moments $\phi = \phi_1 - \phi_2$. The four states are indicated by vertical solid (i, iii) and dashed (ii, iv) lines. }
    \label{fig:four-states}
\end{figure}

We have identified the possibility of simultaneously switching the $p$-wave spin-polarized and ferroelectric order in $\mathrm{GdMn_2O_5}$. Such a possibility is not necessarily present in all ferroelectric $p$-wave materials. We first discuss the states with reversed order parameters and the symmetry operations connecting them, as shown in \autoref{fig:four-states}. Different sets of symmetry operations reverse the $p$-wave and the ferroelectric orders. For example, $[M_y||C_{2z}]$ reverses the ferroelectric polarization, while leaving the $p$-wave spin order intact, as illustrated on states (i) and (ii) in \autoref{fig:four-states}(a). In contrast, $[M_y || E]$ switches the $p$-wave order without changing the ferroelectric polarization, see states (ii) and (iii) in \autoref{fig:four-states}(a).

Remarkably, all four states can be described within the same crystal unit cell, only changing the magnetic moment angles $\phi_1$, $\phi_2$ indicated in~\autoref{fig:BiFeO-GdMnO}(d). The spin switching operation, $[M_y||E]$, transforms the characteristic angles as $\phi_1 \to -\phi_1$, $\phi_2 \to -\phi_2$. The spontaneous electric polarization reversing symmetry, $[M_y||C_{2z}]$, acts as $\phi_1 \to -\phi_1+180^{\circ}$, $\phi_2 \to -\phi_2$. Taken together, both order parameters can be switched via $[E||C_{2z}]$ which acts as $\phi_1 \to \phi_1+180^{\circ}$ and $\phi_2 \to \phi_2$ as seen in \autoref{fig:four-states}(a). Remarkably, the states (i) and (iii) are exactly the states involved in the switching by magnetic field in experiments on GdMn\textsubscript{2}O\textsubscript{5}~\cite{ponet_topologically_2022}. 

We now discuss the energetics of the four states in terms of their symmetries.
All four states are energetically degenerate when a nonrelativistic Hamiltonian is considered, as the total energy is invariant under the symmetry operations connecting the four states among themselves. However, this ceases to be true after taking SOC into account, which splits the four-state degeneracy into two pairs of energetically degenerate states. This can be explained in terms of symmetries.
A spin symmetry in the form $[R || R]$ corresponds to a magnetic symmetry operation and thus conserves the energy coming from the relativistic Hamiltonian. Here, only the diagonal symmetry operation $[E || C_{2z}]$ can be modified into such a form by applying symmetry operations present in the material: $[E || C_{2z}] [C_{2z} || E] = [C_{2z} || C_{2z}]$. The experimental states (i) and (iii)~\cite{ponet_topologically_2022} are therefore the degenerate ground states, while states (ii) and (iv) are degenerate excited states.
Remarkably, the new ground states either have electric and spin polarizations which are both positive, or both negative. The signs of the two polarizations are therefore coupled, and the phenomenon we dub $p$-wave-magnetoelectric coupling emerges. Consequently, by externally acting on one ferroic order, the other ferroic order can be switched. A concrete example of this would be the switching of the $p$-wave-magnetic order via external electric or magnetic field, as was demonstrated experimentally~\cite{ponet_topologically_2022}.

One can take advantage of the fact that the four states share the same parent structure to study a hypothetical switching path between them that only involves general variation of $\phi_1$, $\phi_2$, i.e., rotation of magnetic moments, such that the coplanar $p$-wave order is maintained. In this work, we study a switching path along the four states in the order (i-ii-iii-iv-i) via linear variation of $\phi_1$ and $\phi_2$, such that $\phi_1$ increases and makes a full revolution, while $\phi_2$ oscillates between $\pm\phi_0$.

Similar switching paths, based solely on varying $\phi_1$ and $\phi_2$ have been suggested in previous works about magnetoelectric switching in $\mathrm{GdMn_2O_5}$~\cite{ponet_topologically_2022,PhysRevLett.134.016708}. Similarly to  our work, $\phi_1$ makes a full revolution during one switching round, while $\phi_2$ oscillates around a fixed angle. Our switching path and its comparison to the previously suggested paths is discussed in the Supplemental Material in greater detail~\cite{supp}.

\autoref{fig:four-states}(b) shows the calculated dependence of characteristic parameters of the two ferroic orderings -- electric polarization $P$ and spin splitting strength $\Delta$ -- versus relative angle of magnetic moments $\phi = \phi_1 - \phi_2$. Without SOC, both dependencies are sinusoidal with a relative phase shift of $\pi/2$ between them, i.e., spin splitting is zero when electric polarization is the largest and vice versa. Without SOC, the spin and orbital spaces are decoupled, and the relevant observables only depend on the relative angle between the magnetic moments $\phi = \phi_1 - \phi_2$, not the absolute angles $\phi_1$ and $\phi_2$ similarly to the previous studies.
Consequently, the nonrelativistic angular dependence of $P$ and $\Delta$ along our switching path, presented in \autoref{fig:four-states}(b), also holds for the previously suggested paths~\cite{ponet_topologically_2022,PhysRevLett.134.016708}, even though the paths are slightly different, because they share identical $\phi$ evolution.

The spin splitting strength $\Delta$, shown in~\autoref{fig:four-states}(b), further shows that the spin splitting originates from the noncollinearity of the magnetic moments: the collinear magnetic order (at $\phi_1 = \phi_2 = 0°$) reduces our system from a $p$-wave magnet to a collinear antiferromagnet without nonrelativistic spin splitting~\cite{smejkal_beyond_2022}.

The relativistic symmetry breaking visibly shifts the electric polarization dependence, with the maximum shifting away from the collinear case.
The calculated electric polarization maximum is $\Delta P = 0.04$~e\AA~per unit cell, which results in a macroscopic value of $P = 0.09~\mathrm{\mu C~cm^{-2}}$. The experimentally reported value by Ponet \textit{et al.}~\cite{ponet_topologically_2022} is $\sim 0.15~\mathrm{\mu C~cm^{-2}}$. The mismatch between the theoretical and experimental value can be attributed to limited accuracy of the first-principles method and additional ionic contributions, which have not been considered here.

\textit{Discussion.}--- 
Finally, we discuss a memory application based on insulating ferroelectric $p$-wave magnets. Such a memory device can be built from a double-layer structure in which a metal with a fixed $p$-wave order serves as the reference layer, while a ferroelectric insulating $p$-wave magnet acts as the switchable free layer. A similar device structure has previously been considered with altermagnets~\cite{PhysRevLett.134.106802}, but requires different writing mechanisms, such as the altermagnetoelectric effect~\cite{smejkal2024altermagneticmultiferroicsaltermagnetoelectriceffect}. Our device structure, with only two components (an insulating and a metallic $p$-wave magnet), is in principle simpler than the originally proposed altermagnetic tunneling junctions with three components (two altermagnetic metallic leads separated by a nonmagnetic tunneling barrier)~\cite{PhysRevX.12.011028, shao_spin-neutral_2021}.
Logical states in our device correspond to the parallel or antiparallel alignment of the $p$-wave spin polarization in the metallic and insulating layers. The parallel or antiparallel state can be read out through a large spin-filtering effect
previously discussed for a different geometry with altermagnets~\cite{samanta_spin_2025}.

\textit{Conclusion.}---
We introduce ferroelectric odd-parity-wave magnets and classify them into Type-I, Type-IIa, and Type-IIb categories based on the origin of the polar ordering, supported by spin and magnetic symmetry group theory. Furthermore, we present 52 experimentally available material candidates, with the majority in the Type-II category. This starkly contrasts the 44 ferroelectric altermagnetic candidates, which overwhelmingly belong to Type-I~\cite{smejkal2024altermagneticmultiferroicsaltermagnetoelectriceffect}.

We investigate the interplay of the $p$-wave-magnetic and ferroelectric order more closely in Type-IIa multiferroic material candidate $\mathrm{GdMn_2O_5}$. While the ferroelectric switching in $\mathrm{GdMn_2O_5}$ has already been demonstrated~\cite{ponet_topologically_2022, PhysRevLett.134.016708}, we show that (I) $\mathrm{GdMn_2O_5}$ is not a conventional antiferromagnet, but rather an unconventional $p$-wave magnet, and (II) the ferroelectric switching is accompanied by a simultaneous reversal of the $p$-wave order. 

\textit{Acknowledgments.}---
We thank Johannes Mitscherling and Clara Geschner for helpful discussions.
We acknowledge funding from the ERC Starting Grant No. 101165122 and the ERC Advanced Grant no. 101095925.  This work was supported by the Ministry of Education, Youth and Sports of the Czech Republic through the e-INFRA CZ (ID:90254).



%

\end{document}